\documentclass[aps,superscriptaddress,preprintnumbers,twocolumn,prl,showpacs]{revtex4-1}
\usepackage{graphicx}
\usepackage{dcolumn}
\usepackage{bm}
\usepackage{epsfig}
\usepackage{color}

\usepackage{longtable}
\usepackage{amsmath}
\usepackage{multirow}

\begin{document}


\title{Surface reconstruction, premelting, and collapse of open-cell nanoporous Cu via thermal annealing}

\author{L.~Wang}
\affiliation{College of Science, Hunan Agricultural University, Changsha 410128, People's Republic of China}

\author{X. M. Zhang}
\affiliation{College of Science, Hunan Agricultural University, Changsha 410128, People's Republic of China}

\author{L.~Deng}
\email{Corresponding author. dengl@hunau.edu.cn}
\affiliation{College of Science, Hunan Agricultural University, Changsha 410128, People's Republic of China}

\author{J. F.~Tang}
\email{Corresponding author. jftang@hunau.edu.cn}
\affiliation{College of Science, Hunan Agricultural University, Changsha 410128, People's Republic of China}

\author{S. F. Xiao}
\affiliation{Department of Applied Physics, Hunan University, Changsha 410001, People's Republic of China}

\author{H. Q. Deng}
\affiliation{Department of Applied Physics, Hunan University, Changsha 410001, People's Republic of China}

\author{W. Y. Hu}
\affiliation{College of Material Science and Engineering, Hunan University, Changsha 410001, People's Republic of China}

\date{\today}

\begin{abstract}
We systematic investigate the collapse of a set of open-cell nanoporous Cu (np-Cu) with the same porosity and shapes, but different specific surface area, during thermal annealing, via performing large-scale molecular dynamics simulations. Surface premelting is dominated in their collapses, and surface premelting temperatures reduce linearly with the increase of specific surface area. The collapse mechanisms are different for np-Cu with different specific surface area. If the specific surface area less than a critical value ($\sim$ 2.38 nm$^{-1}$), direct surface premelting, giving rise to the transition of ligaments from solid to liquid states, is the cause to facilitate falling-down of np-Cu during thermal annealing. While surface premelting and following recrystallization, accelerating the sloughing of ligaments and annihilation of pores, is the other mechanism, as exceeding the critical specific surface area. The recrystallization occurs at the temperatures below supercooling, where liquid is instable and instantaneous. Thermal-induced surface reconstruction prompts surface premelting via facilitating local ``disordering'' and ``chaotic'' at the surface, which are the preferred sites for surface premelting.
\end{abstract}

\maketitle
\section{I. Introduction}
Nanoporous metals, with open-cell network structures, are comprised of interconnected ligaments of nanometer-scale characteristic dimenstion, resulting in a large surface area. Their stabilities and morphological properties determine the promising applications as sensors\cite{hu08ac}, actuators \cite{biener09nm}, battery electrodes \cite{snyder10nm}, catalysts \cite{erlebacher01nat}, and in biomedicine \cite{swan05bm}. During thermal annealing, microstructural changes, including heating-induced void collapse and the reduction in volume can be observed by implementing the atomic-scale simulations \cite{crowson07sm, crowson09sm, kolluri11am} and experiments using electron microscopy and X-ray diffractions \cite{parida06prl, kucheyev06apl, seker07am, jin09am, chen12am, frenkel12prb}. Surface melting, via accelerating the diffusion of atoms in nanoporous metals, is a dominant factor to prompt the occurrence of the pinch-off in ligament, the smooth-out for surface curvature, and the formation of enclosed voids in ligaments \cite{hakamada09jmr, chen10apl}. Consequently, significant insights of the phenomenology and underlying mechanisms of surface melting in metallic nanoporous, such as the relation between surface configuration and microstructure evolution, are key issues.

Surface melting was believed to be impossible to superheat a crystal above its equilibrium melting point, indicating a premelting \cite{cahn86nat, lipowsky89prl, curtin89prb}, which has been disproved by the subsequent studies on melting of instable nanoparticles \cite{lai96prl, guisbiers09jpcc, yang08jpcc, zhu09jpcc, qi04mcp, qi05pb, qi16acr} and other nanostructures \cite{shin07apl, li08jap} in experiments and theories. Surface premelting is also observed in nanoporous materials, and melting temperature was related to the size of pores. A roughly proportional function is shown between the melting temperature and reciprocal of averaged pore-size, as predicted by Gibbs-Thomson equation \cite{evans87jcp}, which is consistent with the experimental reports using nuclear magnetic resonance (NMR) \cite{strange98mp} and X-ray diffraction techniques \cite{morishige97jcp} and the atomistic simulation studies \cite{miyahara97jcp, gubbins14jct, coasne13csr, coasne15lm}. These studies enriched our knowledge of the melting in nanoporous materials. However, the thermal-induced variety of surface microstructure, its effect on the melting, and the collapse processes of nanoporous remain to be revealed.

In this work, we implement a systematic study on the melting behavior in open-cell nanoporous Cu (np-Cu), with the same porosity and shape approximately, and different specific surface areas by performing large-scale molecular dynamics (MDs) simulations. Cu is chosen owing to its accordance between experiments and simulations \cite{japel05prl, trautt12am, zheng07jcp}, as well as its accurate embedded-atom method potential \cite{mishin01prb}, which well describes the melting of Cu \cite{zheng07jcp}. Our simulations reveal that the collapse of np-Cu is prompted via surface melting, or melting and following recrystallization, which is dependent on the specific surface area. In particular, our results show that surface prefers to melt at the disordering sites as the result of surface reconstruction as the temperature elevation. The MD methodology are presented in Sec.II, and results and discussion in Sec. III, followed by summary and conclusions in Sec.IV.

\begin{table*}
\centering
\caption{\label{config}Characteristics and surface premelting temperatures for two groups of np-Cu samples with different system sizes. $\rho_r$: relative mass density; $A_{\rm solid}$: surface area; $V_{\rm solid}$: solid volume; $\gamma$: specific surface area; $T_{\rm C}$: collapse temperature of np-Cu; $T_{\rm M}$: equilibrium melting temperature for defect-free bulk Cu.}
\begin{tabular}{cccccccccc}
\hline \hline
   Group & $\#$ & $k$ & Size ($\times$ 10$^6$) & $\rho_r$ ($\%$) & $A_{\rm solid}$ ($\times$ 10$^{4}$ nm$^2$) & $V_{\rm solid}$ ($\times$ 10$^{4}$ nm$^3$) & $\gamma$ (nm$^{-1}$) & $T_{\rm C}/T_{\rm M}$ & Dimension (nm$^3$) \\
\hline
 \multirow{10}{*}{Group $I$}& $A1$ &  $1$ & 2.33 & 58.19 & 0.45 & 2.44 & 0.18 & 1.00 & \multirow{10}{*}{36.15$\times36.15\times$36.15}\\
  & $A2$ &  2 & 2.33 & 58.22 & 0.91 & 2.45 & 0.37 & 0.96 & \\
  & $A3$ &  4 & 2.32 & 58.11 & 1.89 & 2.42 & 0.78 & 0.93 & \\
  & $A4$ &  6 & 2.33 & 58.22 & 2.92 & 2.43 & 1.20 & 0.89 & \\
  & $A5$ &  8 & 2.32 & 58.07 & 4.08 & 2.45 & 1.67 & 0.81 & \\
  & $A6$ & 10 & 2.35 & 58.70 & 5.11 & 2.40 & 2.13 & 0.70 & \\
  & $A7$ & 11 & 2.33 & 58.19 & 5.77 & 2.40 & 2.38 & 0.67 & \\
  & $A8$ & 12 & 2.33 & 58.18 & 6.36 & 2.42 & 2.63 & 0.59 &\\
  & $A9$ & 13 & 2.33 & 58.19 & 6.99 & 2.40 & 2.94 & 0.52 &\\
 & $A10$ & 14 & 2.33 & 58.22 & 7.58 & 2.42 & 3.13 & 0.48 &\\
\hline
 \multirow{7}{*}{Group $II$}& $B1$  &  1 & 0.29 & 58.22 & 0.12 & 0.32 & 0.37 & 0.96 & \multirow{7}{*}{18.08$\times$18.08$\times$18.08}\\
  & $B2$  &  2 & 0.29 & 58.68 & 0.25 & 0.32 & 0.77 & 0.93 & \\
  & $B3$  &  3 & 0.29 & 58.22 & 0.38 & 0.31 & 1.20 & 0.81 & \\
  & $B4$  &  4 & 0.29 & 58.07 & 0.51 & 0.31 & 1.67 & 0.78 & \\
  & $B5$  &  5 & 0.29 & 58.70 & 0.64 & 0.30 & 2.13 & 0.67 & \\
  & $B6$  &  6 & 0.29 & 58.11 & 0.80 & 0.30 & 2.67 & 0.59 & \\
  & $B7$  &  7 & 0.29 & 58.22 & 0.95 & 0.30 & 3.14 & 0.52 & \\
\hline\hline
\end{tabular}
\end{table*}

\begin{figure*}[t]
\includegraphics[scale=0.8]{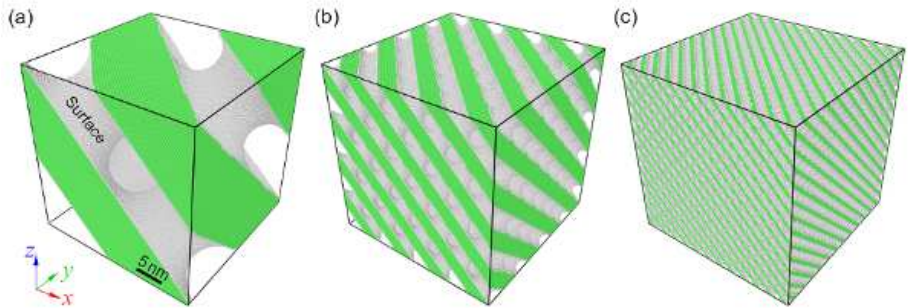}
\caption{\label{npmodel} Atomic configurations of relaxed single-crystal nanoporous Cu (np-Cu) with identical relative mass density ($\rho_r \approx 58.20\%$), but different specific surface area $\gamma \approx 0.18$ ($\#A1$, a), 1.20 ($\#A4$, b), and 2.63 nm$^{-1}$ ($\#A8$, c), respectively, with color coding based on common neighbor analysis. FCC atoms and surface atoms are colored with green and gray, respectively.}
\end{figure*}

\section{II. Methodology}
We utilize the Large scale Atompic/Molecular Massively Parallel Simulator (LAMMPS) \cite{lammps} and an accurate embedded atom method potential, proposed by Mishin \cite{mishin01prb}, describing the atomic interaction in Cu, for our MD simulations. The potential has been fitted to reproduce physical properties such as stacking fault energy and elastic moduli, and widely used in lots of MD simulations, inculding melting \cite{li15jcp, wang16prb}, equation of state \cite{bringa04jap}, and plastic deformation \cite{bringa05sci}.

To construct the MD models, we first use LAMMPS to generate a single crystal cubic bulk oriented in the $\langle100\rangle$ directions, and the dimensions are $\sim 36.15 \times 36.15 \times 36.15$ nm$^3$. Then the diamond-liked open-cell np-Cu \cite{cui17jap} is obtained by removing atoms from the bulk crystal, which obey
\begin{equation}
\begin{aligned}
     \mu \ge & \sin{\left(2\pi x \cdot k\right)}\sin{\left(2\pi y\cdot k\right)}\sin{\left(2\pi z\cdot k\right)}+\\
     &\sin{\left(2\pi x\cdot k\right)}\cos{\left(2\pi y\cdot k\right)}\cos{\left(2\pi z\cdot k\right)}+ \\
     &\cos{\left(2\pi x\cdot k\right)}\sin{\left(2\pi y\cdot k\right)}\cos{\left(2\pi z\cdot k\right)}+ \\
     &\cos{\left(2\pi x\cdot k\right)}\cos{\left(2\pi y\cdot k\right)}\sin{\left(2\pi z\cdot k\right)},
\end{aligned}
\end{equation}
where $x$, $y$, and $z$ are the fractional coordination of removed atoms, $\mu$ (-2.0 $\le$ $\mu$ $\le$ 2.0) and $k$ are two key factors to control the shape of nanopores in bulk, relative mass density ($\rho_r$, the density ratio of the np-Cu with respect to full density Cu), and specific surface area ($\gamma$), defined as surface area ($A$) to solid volume ($V_{\rm solid}$). The specific surface area of the np-Cu structure is computed with surface analysis \cite{ovito}. Increasing $k$ and keeping $\mu$ constant, $\mu$=-0.2,, the shape and $\rho_r$ almost keep invariant, $\sim$0.582, and $\gamma$ increases as the ligaments narrow. We here construct 10 np-Cu with different specific surface area, and three representative configurations are shown in Fig.~\ref{npmodel}. The similar np-Cu with the smaller system size are also attempted to check possible size effects, which comprise about 0.29 million atoms with dimensions of $\sim$ 18.08 $\times$ 18.08 $\times$ 18.08 nm$^{3}$. The details of all np-Cu samples are summarized in Table~\ref{config}.

The configurations are first relaxed with the conjugate gradient method, and then melting simulations are performed under three-dimensional periodic boundary conditions with the constant-pressure-temperature ensemble. Temperature, $T$, is controlled with a Hoover therostat, and the isotropic pressure, with isotropic volume scaling. The solids undergo incremental heating (300--1800 K) into the liquid regime at ambient pressure, and the temperature increment is 50 K per 100 ps at high temperatures. The time step for integrating the equation of motion is 1 fs. At each temperature, the run duration is 100 ps.

The local structure around an atom is characterized with the common neighbor analysis \cite{faken94cms,tsuzuki07cpc}, based on which atoms are classified into face-centered cubic (FCC), hexagonal close-packed (HCP), or unknown types. Order parameter, $\psi$, introduced by Morris and Song \cite{morris03jcp} is utilized to identify atoms in liquid and solid structures, which has been successfully used to describe the micromorphological and dynamics properties of melting under both equilibrium \cite{zheng07jcp, li15jcp} and non-equilibrium conditions \cite{he13jcp}. For each an atom, it is defined as
\begin{equation}
 \psi=\left|\frac{1}{N_q}\frac{1}{N_c}\sum_{\bf r}\sum_{\bf q}\exp(i{\bf q} \cdot {\bf r})\right|^2,
\end{equation}
where $N_c$ is the coordination number and vector $\bf{r}$ refers to the distance between the atom and its nearest neighbors. The set of $N_q$ direction vector ${\bf q}$ satisfying $\exp(i{\bf q}\cdot{\bf r})$=1 is chosen for any vector ${\bf r}$ connecting nearest neighbors in perfect fcc solid. The average local order parameter, averaged among the atom and its nearest neighbors, is adopted to better describe local disordering. The global order parameter,
\begin{equation}
\Psi = \frac{1}{N} \sum_i \psi_i,
\end{equation}
which is the average of $\psi$ over all $N$ atoms in the system, is also used in the following discussion. 

To definitively distinguish a melt (liquid) from crystalline and amorphous solids, self-diffusion coefficient, $D$, is calculated with Einstein expression \cite{rapaport95b},
\begin{equation}
 D = \lim_{t\to\infty}\frac{1}{6t}{\rm MSD}(t),
\end{equation}
from mean square displacement MSD,
\begin{equation}
 {\rm MSD}(t) = \langle\left|r(t)-r(0)\right|^2\rangle.
\end{equation}
Here $t$ denotes time, $r$ is the atomic position, and $\langle\cdots\rangle$ denotes averaging over ensemble only.

\begin{figure}
\includegraphics[scale=0.59]{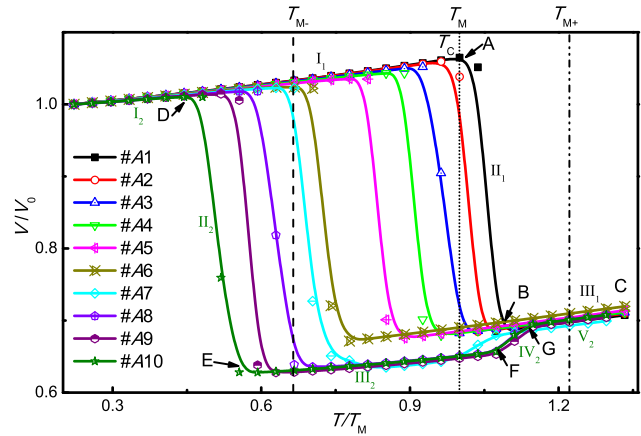}
\caption{\label{mc} Volume ($V/V_0$) versus temperature ($T/T_{\rm M}$) for Cu nanoporous during incremental heating. Different symbols denote np-Cu with different $\gamma$ ($\#A1-A10$, Table I). A (or D): the commence of collapse; B (or E): the finish of collapse; F and G: the transition points for accelerating and decelerating volume expansions, respectively. The dash, dotted and dash-dotted lines denote the limit temperature of supercooling ($T_{\rm M-}$), the equilibrium melting temperature ($T_{\rm M}$), and the limit temperature of superheating ($T_{\rm M+}$), for defect-free bulk Cu, respectively.}
\end{figure}

\begin{figure*}
\includegraphics[scale=1.0]{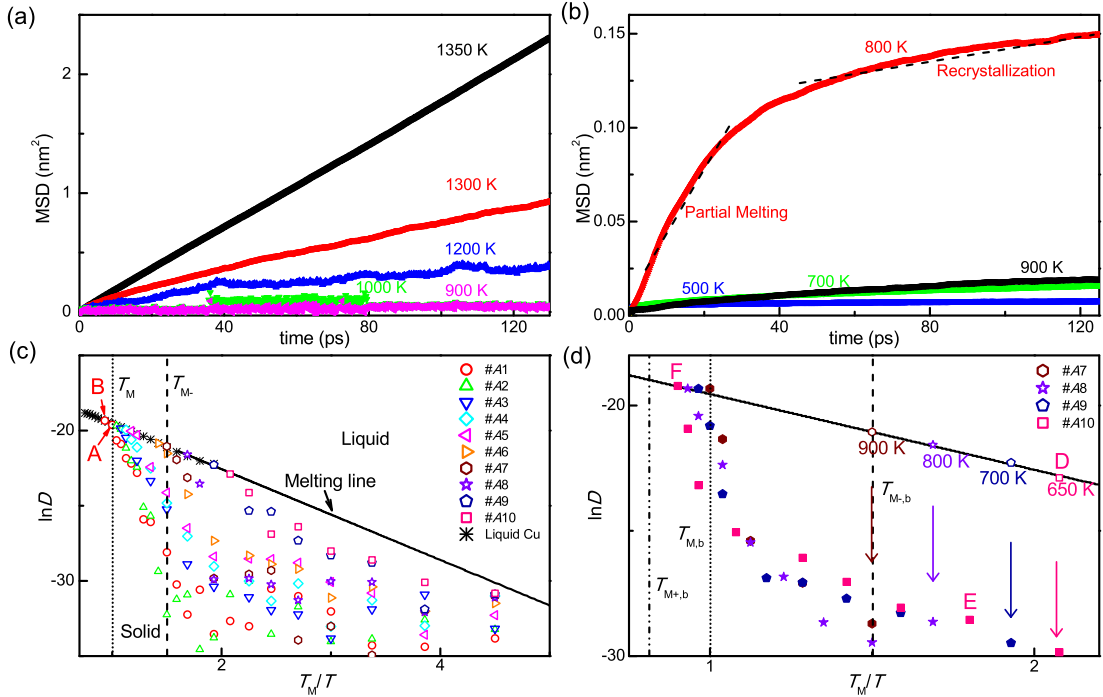}
\caption{\label{dif}(a) and (b) Time evolution of MSDs in the surface region of np-Cu $\#A$1 and $\#A$8 at different temperature, respectively.  (c) Diffusion coefficients as a function of temperature, of liquid Cu ($*$), and in the surface region of np-Cu$\#A1-A6$ and $\#A7-A10$ before the collapse ($T \le T_{\rm C}$). The solid line is the Arrhenius plot for Liquid Cu at zero pressure. (d) Relations of diffusion coefficients and temperature of the np-Cu $\#A$7--$\#A$10 during the surface premelting (empty symbols), crystallization and system melting (filled symbols).}
\end{figure*}

\section{III. Results and discussion}
\subsection{Collapse of np-Cu during thermal annealing}
Our simulations present similar melting phenomena for the np-Cu with different system size, that is the system size effect can be neglected for this work. We here use np-Cu samples with the larger system ($\sim$36.15$\times36.15\times36.15$ nm$^3$, Group $I$) for the following discussion. The initial system-averaged volume $V_0$,  at 300 K, for all nanoporous, are about 4.66$\times$10$^4$ nm$^3$. We here first clarify the definition of three temperatures, $T_{\rm M}$, $T_{\rm M-}$ and $T_{\rm M+}$, used in this work. $T_{\rm M}$ is defined as the equilibrium melting temperature of a defect-free bulk Cu system without surface or other defects, which is about 1350 K \cite{samanta14sci}, agreement with that in experiment ($\sim$1356 K) \cite{brillo03ijt}. $T_{\rm M-}$ and $T_{\rm M+}$ are the limit temperature of supercooling liquid and superheating solids under equilibrium conditions, which are about 900 K and 1650 K for Cu, respectively.

Figure~\ref{mc} shows the temperature evolutions of system-averaged volume, $V$, for 10 np-Cu samples ($\#A1-A10$). With the increase of temperature, $V/V_0$ undergoes a linear increase nearly at first (I$_1$ and I$_2$) until an abrupt drop in the slope (II$_1$ and II$_2$), owing to the ``onset'' of collapse for all the nanoporous Cu. The np-Cu collapse could be resulted from coarsening of nanopores \cite{kolluri11am} or premelting \cite{zheng07jcp, li15jcp} from the surface, since temperature lies well below the equilbrium melting temperature $T_{\rm M}$ \cite{jin01prl, forsblom05nm}. The temperatures for collapse, $T_{\rm C}$, are observed to decrease from 1350 K ($T/T_{\rm M}=1.0$, D for $\#A1$) to 650 K ($T/T_{\rm M}=0.48$, A for $\#A10$) as the increase of specific surface area, $\gamma$ (see TABLE~\ref{config}).

Furthering to increase temperatures, two different evolution modes are shown for the nanoporous Cu with different $\gamma$, and a critical specific surface area $\gamma_{\rm crit}$, is about 2.38 nm$^{-1}$. For np-Cu with $\gamma < \gamma_{\rm crit}$ ($\#A1-A6$), a linear $V/V_0$ rising stage (III$_1$) follows the completion of collapse (B), and their $V-T$ functions are almost same with each other. For np-Cu with $\gamma \geq \gamma_{\rm crit}$ ($\#A7-A10$), $V/V_0$ first increases linearly (III$_2$) after collapse completing (E), which is below the $T-V$ functions at stage III$_1$ for np-Cu $\#A1-A6$. Then an abrupt increase stage of $V/V_0$ (IV$_2$) is observed as $T$/$T_{\rm M}\geq$ 1, followed by a stage of V$_2$ as $V/V_0$ increasing linearly and slowly, which shows the similar processes with solid melting \cite{zheng07jcp, li15jcp}. It is believed to be solid phases for the np-Cu samples at the stage of III$_2$, while partial melting at IV$_2$ and completed liquid phases at V$_2$ stages. Interestingly, $V-T$ functions at III$_1$ and V$_2$ are observed to well coincide with each other, implying the completed melting during the stage III$_1$. It is deduced that the melting of nanoporous dominates the collapse (II$_1$) for np-Cu $\#A1-A6$. During the nanoporous melting, surface premelting should be preferred owing to its instability and higher energy \cite{cahn86nat}. However, surface melting is inadequate to describe the collapse (II$_2$) in np-Cu $\#A7-A10$, owing to the solid stage of III$_2$. We here assume a melting and recrystallization process occurs to accelerate the collapse of nanoporous Cu at II$_2$ stage.

To identify the surface premelting and recrystallization processes in Cu nanoporous, self-diffusion coefficients, $D$, for the surface atoms (the disordering atoms in Fig.~\ref{npmodel}), are calculated from MSD at different heating temperatures. Here, two MSD evolutions during the extended equilibration period at different temperatures, for np-Cu $\#A1$ and $\#A8$, are taken as the examples shown in Fig.~\ref{dif} (a) and (b), respectively. For np-Cu $\#A1$, MSDs($t$) increase linearly with different slopes, $D$, at different temperature. The calculated $D$ increases from 2.13 $\times$ 10$^{-11}$ to 2.95 $\times$ 10$^{-9}$ m$^2$ s$^{-1}$, as $T$ increases from 900 to 1350 K. However, MSD($t$) exhibits two distinct stages for np-Cu $\#A8$ at 800 K: rapid increase during the first 0--40 ps ($D\approx 3.58 \times 10^{-10}$ m$^{2}$ s$^{-1}$), which is attributed to a partial melting of surface, and followed by slower increase ($D \approx 3.64 \times 10^{-13}$ m$^2$ s$^{-1}$), owing to the crystallization of supercooled melts. In contrast, MSD($t$), at $T$ = 500, 700 and 900 K, remain approximately constant, with $D \sim 10^{-13}$ m$^2$ s$^{-1}$. The difference of orders of magnitude in $D$ between different temperatures supports the partial melting of surface region for different Cu nanoporous.

\begin{figure}
\includegraphics[scale=0.55]{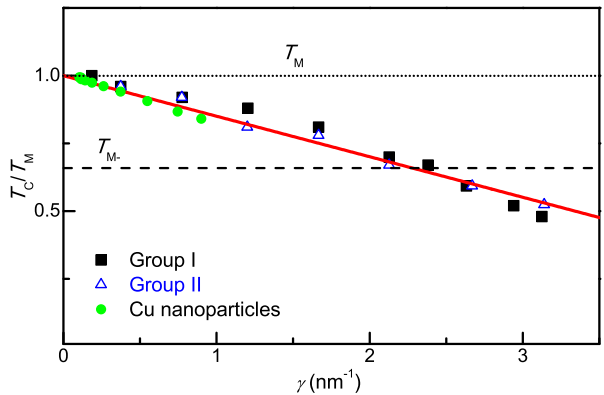}
\caption{\label{scm}The function of temperature for np-Cu collapse, $T_{\rm C}/T_{\rm M}$, of nanoporous Cu (Group I and II) and Cu nanoparticles \cite{zhu09jpcc} in the simulations, and specific surface area, $\gamma$. The red solid line denotes the function of $T_{\rm M}$ and $\gamma$ according to Eq.~\ref{fd3}, respectively. The dotted and dash lines denote the limit temperature of supercooling liquid and equilibrium melting temperature, respectively. }
\end{figure}

To further verify the surface melting, we then compare diffusion coefficients at different temperatures with those of liquid copper at zero pressure [Fig.~\ref{dif}(c) and (d)]. Temperature is varied for liquid copper in order to obtain an Arrhenius plot, and its extention to $T_{\rm M}/T > 1$ means supercooled liquid, which can be considered as the melting line. Under/in melting line, it represents a liquid; otherwise a solid. At low temperatures, ln$D$ for np-Cu $\#A1-A10$ fluctuate around a constant value, $\sim$10$^{-14}$ m$^2$ s$^{-1}$, but follow an approximately linear increase at high temperatures owing to the acceleration of surface diffusion. As $T=T_{\rm C}$, ln$D$ at $T_{\rm C}$ are located in the melting line, implying the commence of surface melting [Fig.~\ref{dif}(c)]. Further to increase temperatures ($T>T_{\rm C}$), ln$D$ for np-Cu $\#A1-A6$ are also observed to lie in melting line, supporting the deduction above in which surface melting promotes the collapse of surface (II$_1$ in Fig.~\ref{mc}). However, ln$D$ for np-Cu $A7-A10$ are observed to decrease abruptly at $T=T_{\rm C}$ [Fig.~\ref{dif}(d)], which means the recrystallization of liquid. Then it can be concluded that surface melting and recrystallization is indeed dominated in the np-Cu collapse for thermal-annealed np-Cu $\#A7-A10$. Interestingly, their $T_{\rm C} \leq T_{\rm M-}$, suggesting the instable liquid phase.  During stage III$_2$, ln$D$ always lie below the melting line, and are shown to increase via the incremental heating until Cu samples melt again. The second melting temperatures for Cu nanoporous are observed to be $T \geq T_{\rm M}$, indicating a superheating which occurs in defect-, or surface-free bulk Cu crystals \cite{zheng07jcp}. All the second melting temperature for Cu samples $\#A7-A10$ are $T < T_{\rm M+}$, implying the existence of defects \cite{li15jcp}, and it will be discussed below.


Based on the results above, the relations of surface premelting temperature (or collapse temperature) and specific surface area, $\gamma$, for np-Cu samples are obtained, as shown in Fig.~\ref{scm}. Two sets of np-Cu samples with different system size (Group $I$ and $II$) are considered. An approximately linear decrease of melting temperature ($T_{\rm C}$) of np-Cu is observed as increasing $\gamma$, which is in accordance with the results of Cu nanoparticles \cite{zhu09jpcc}. Interestingly, the effect of system size can be neglected, owing to the same melting temperature for Group $I$ and $II$ np-Cu with the same $\gamma$. Fitting the function of $\gamma$--$T_{\rm C}$ with Eq.~\ref{fd3} in Appendix, it shows a perfect consistence with the results of nanoparticles and nanoporous. Consequently, the melting of nanostructures is dependent strongly on specific surface area, and we can predict their melting temperatures from the function. In addition, it also implies that only surface melting occurs for the nanostructures with melting points $T_{\rm M-} < T \leq T_{\rm M}$ , while a transient melting and following recrystallization predominates for ones with melting points below $T_{\rm M-}$.

\begin{figure}
\includegraphics[scale=0.50]{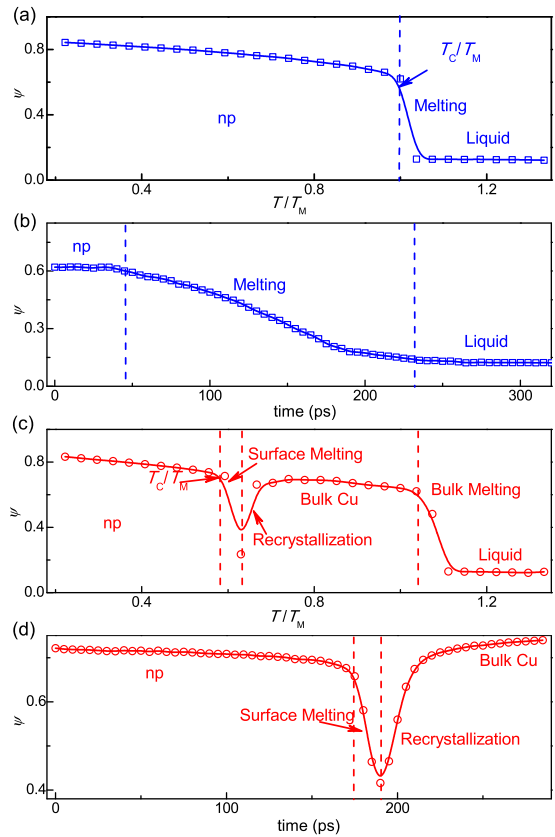}
\caption{\label{op}Global order parameter, $\Psi$, as a function of temperature (a) and (c) and of time at 1350 K (b) and 800 K ($T/T_{\rm M}\approx1$ and 0.59), for np-Cu $\#A$1 (blue) and $\#A$8 (red), respectively.}
\end{figure}

\begin{figure*}
\includegraphics[scale=0.75]{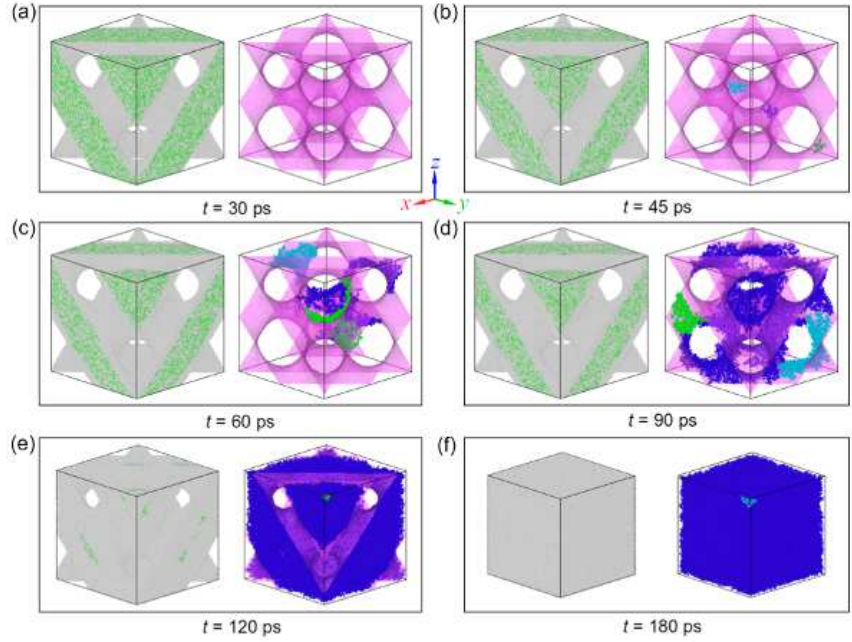}
\caption{\label{n1t}Configurations of $\#A1$ np-Cu colored with CNA and corresponding liquid atoms during melting at 1350 K. Liquid atoms within the first, second, and third largest clusters are color coded as blue, light-blue, and green, respectively. The surface is represented by the pink membrane.}
\end{figure*}

\begin{figure*}
\includegraphics[scale=0.75]{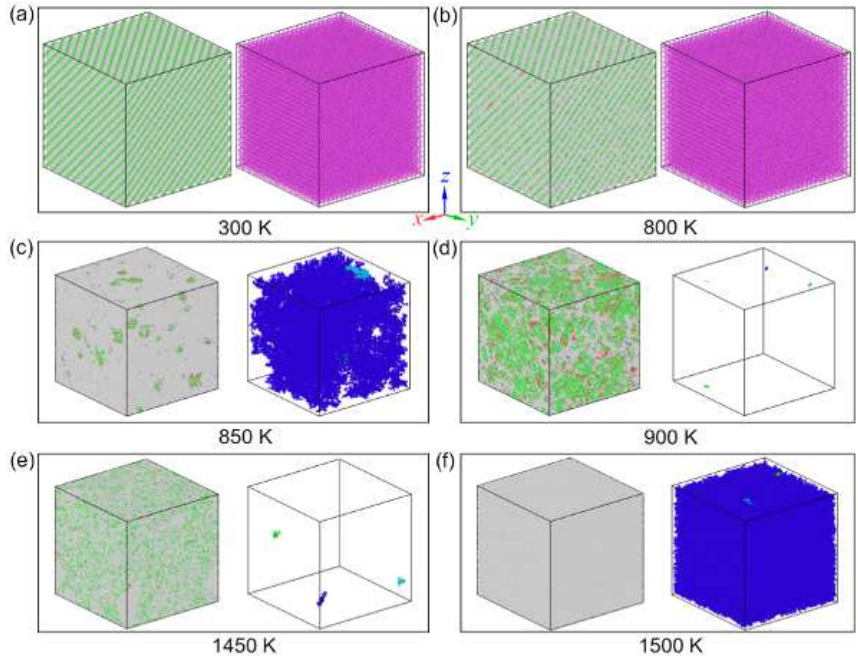}
\caption{\label{n12tm}Configurations of $\#A8$ np-Cu colored with CNA and corresponding liquid atoms during melting during incremental heating. Color coding is same with Fig.~\ref{n1t}.}
\end{figure*}

\begin{figure*}
\includegraphics[scale=0.85]{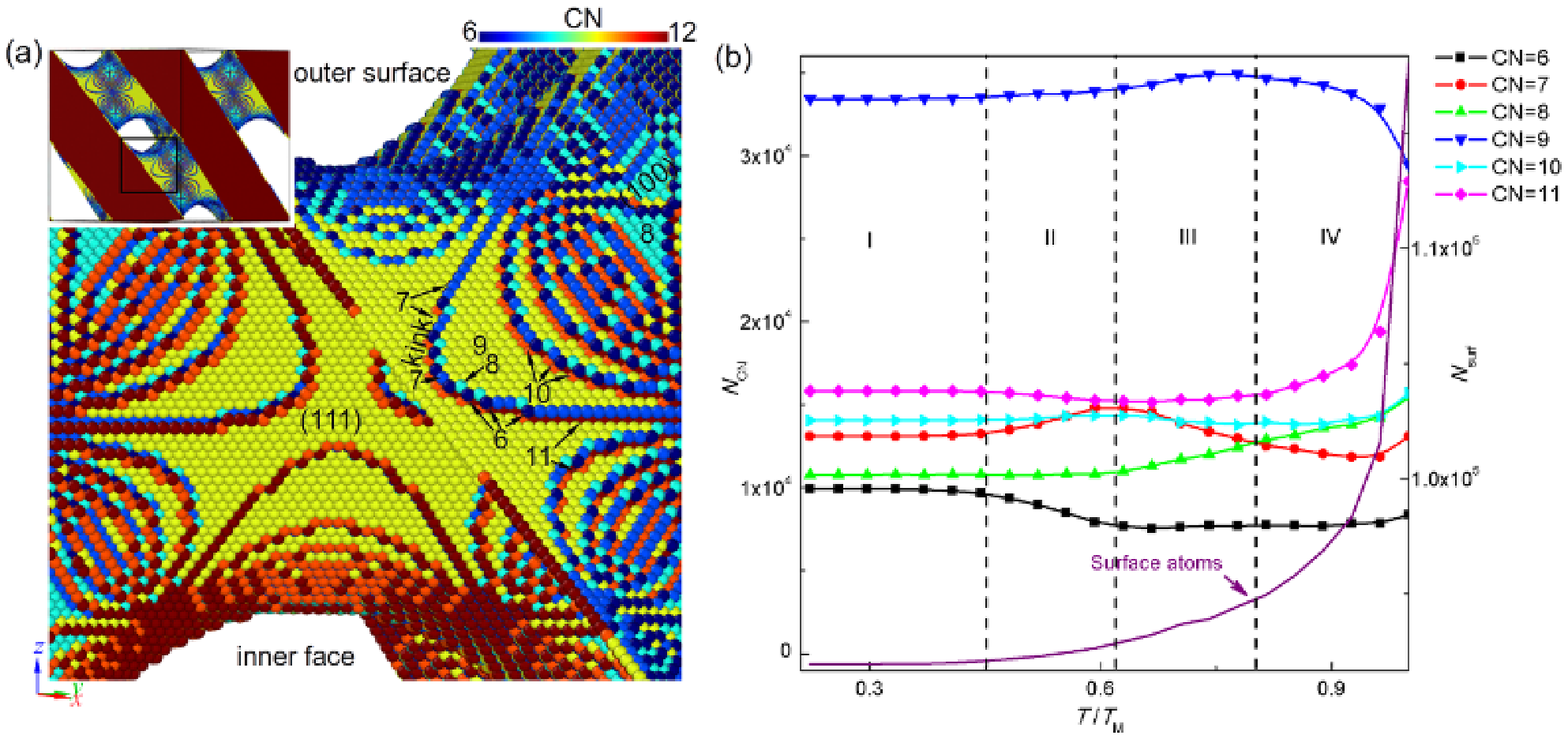}
\caption{\label{cnt}(a) Local surface configuration for np-Cu $\#A1$ (the square region inset) initially, color-coded by coordination analysis. Different coordination number (CN) denoted by a serials of numbers (6--11) represent different surface sites. (b) Evolution of surface atoms with different CN in np-Cu $\#A1$ during incremental heating.}
\end{figure*}

\subsection{Surface melting processes}
To reveal the surface melting of np-Cu, it is necessary to characterize the nucleation and growth process at atomistic scale. The global order parameter, $\Psi$, an important factor to distinguish the liquid and solid phase, is computed to describe the nucleation and growth of melting in nanoporous Cu (Fig.~\ref{op}). For a bulk Cu in liquid stage, $\Psi\approx 0.12$; we thus define an atom as a ``strict liquid'' with high disordering and mobility \cite{levitas12pnas}, if $\psi=0.12$. Liquid atoms tend to aggregate into the clusters, leading to the liquid growth \cite{zheng07jcp}. The cluster analysis \cite{ovito} of liquid atoms is conducted here. When two atoms are within the nearest-neighbor distance with each other, they can be considered as belonging to the same cluster, whose size is defined as the number of liquid atoms within a cluster. Shown in Figs.~\ref{n1t}--\ref{n12tm} are the configuration of liquid clusters during melting. Here, two typical nanoporous ($\#A$1 and $\#A$8) are taken as the examples. Their melting modes are different as their different specific surface area $\gamma_{A1} \approx 0.18 < \gamma_{\rm crit}$ and $\gamma_{A8} \approx 2.63 > \gamma_{\rm crit}$, respectively, corresponding to their different collapse modes discussed above.

For np-Cu $\#A1$, $\Psi$ decreases steadily and slowly at first, and then drops abruptly from 0.62 to 0.12, which means the complete melting of solid in the samples [Fig.~\ref{op}(a)]. The transition temperature is $T\approx$ 1350 K ($T/T_{\rm M}\approx1.0$), consistent with the collapse temperature, $T_{\rm C}$ above (Fig.~\ref{mc}). We also compute $\Psi(t)$ for np-Cu $\#A1$ by furthering heating at 1350 K with the running duration of 320 ps [Fig.~\ref{op}(b)], to describe the evolution process of transition. Continuous heating precipitates the nucleation of surface melting, and $\Psi$ starts to decrease at $t$ $\approx$ 45 ps from 0.62 to 0.12. Sequentially pronounced fluctuations, stable nuclei, and catastropic growth are shown during the process (Fig.~\ref{n1t}). Subcritical nuclei appear at the outer surface via fluctuation in locations and times ($t$ $<$ 60 ps, Fig.~\ref{n1t}), and supercritical nuclei then become stablized and grow at $t \ge 60$ ps. Then liquid clusters grow towards interior grains, giving rise to the void shrink and ligament pinch-off in nanoporous Cu ($t$ = 90 and 120 ps, Fig.~\ref{n1t}). Finally, all voids in np-Cu are observed to be filled by liquid atoms, leading to a completed collapse of surface and the reduction of volume at 180 ps. The melting completes as $t > 230$ ps, with an invariable $\Psi$ approximately, $\Psi\approx$ 0.12.

For np-Cu $\#A8$, interestingly, two decrease stages in $\Psi$ are presented as the incremental heating [Fig.~\ref{op}(c)], different with np-Cu $\#A1$. The first reduction takes place at $T \approx$ 800 K, where $\Psi$ reduces rapidly from 0.72 to 0.24 owing to the nucleation of surface melting, followed by the collapse of nanoporous. At this stage partial atoms in np-Cu transform from solid/quasisolid atoms to liquid/quasiliquid atoms, accelerating the annihilation of pores and collapse. The processes of nucleation and growth for surface melting are shown in Fig.~\ref{n12tm}. These liquid or quasiliquid atoms can be crystallized again as $T < T_{\rm M-}$ K, and $\Psi$ increases from 0.24 to 0.66. The dynamic processes of melting and crystallization can also be described by $\Psi(t)$ [Fig.~\ref{op}(d)]. The surface-free bulk Cu crystal, containing such defects as stacking faults, twins and disordering solids (characterized by CNA), is formed as the completion of crystallization [900 K ($T/T_{\rm M} = 0.67$), Fig.~\ref{n12tm}]. These defects are the sources of melting as heating in further, which accelerates the melting of bulk Cu crystal and reduces its superheating, $T$ = 1450 K ($< T_{\rm M+}$), verifying the assumption above. Then $\Psi$ drops rapidly again from 0.62 to 0.12 as the occurrence of completed melting at 1500 K [$T/T_{\rm M} = 1.11$, Fig.~\ref{n12tm}].

\begin{figure*}
\includegraphics[scale=0.7]{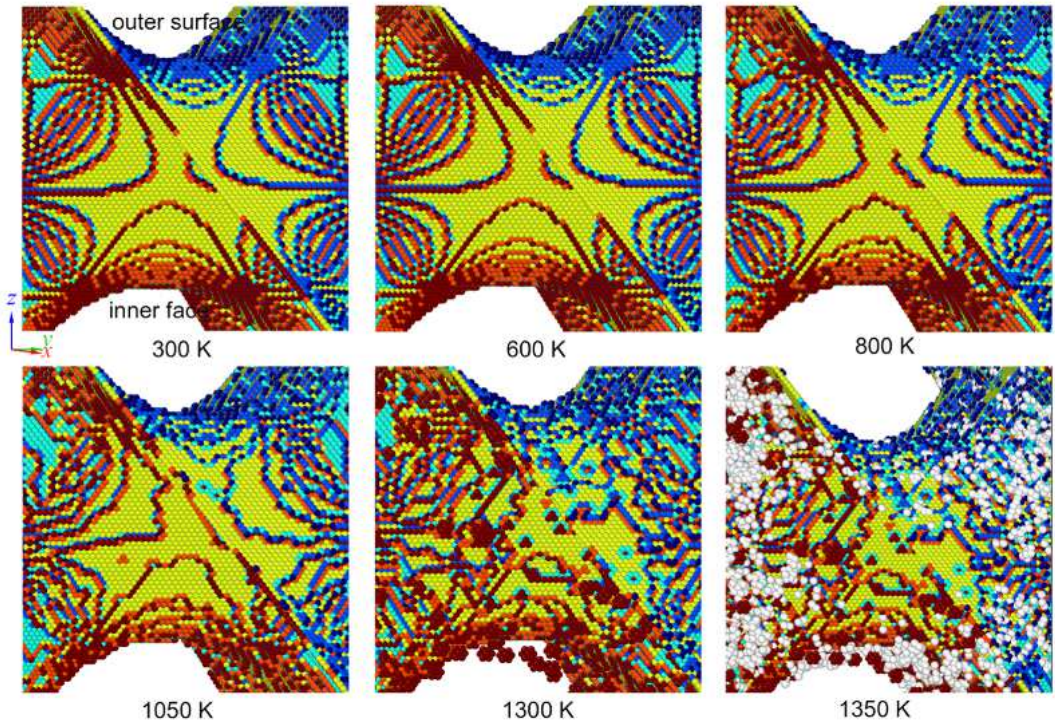}
\caption{\label{sre}The process of surface reconstruction and surface premelting nucleation at 1350 K ($T_{\rm M}$) for np-Cu $\#A$1 during incremental heating. White spheres denote the liquid atoms, and color-coded with CN, same with Fig~\ref{cnt}(a). Only atoms with CN $<$ 12 are displayed.}
\end{figure*}

\subsection{Surface reconstruction}
The phenomena of surface premelting and collapse globally in the np-Cu are presented above, and it shows a strong dependent of specific surface area, $\gamma$. It is known that the melting of defects, such as grain boundaries, is related to microstructure evolution, which is significant to understand the phenomenology and underlying mechanism \cite{alsayed05sci, berry08prb}.  Thus we next examine detailed surface microstructure evolutions via thermal annealing. It also reveals the possible implications and connections to surface premelting. Here np-Cu $\#A1$ is taken as the example as shown in Figs.~\ref{cnt}--\ref{sre}.

Coordination analysis \cite{ovito} are performed after implementing quench, to describe the surface microstructure. The atoms with coordination number, CN $<$ 12, are defined as the surface atoms here, as shown in Fig.~\ref{cnt}(a). In our constructed np-Cu, it is observed that the coordination number of the surface atoms at (100) facet and the close-packed (111) facet are CN$_{\rm s(100)}$ = 8, and CN$_{\rm s(111)}$ = 9 \cite{liu02jacs}, respectively. Several kinked-lines \cite{wang13nm} are formed between two (111) (or 100) facets, where the atoms with CN$_{\rm kl,ed}$ = 7 at the outer surfaces, considered as the edge of facets \cite{deng10jpcc}; while CN$_{\rm kl, sub} = 11$ at the subsurfaces. Between two edges, some kinked points, where the atoms with CN$_{\rm kp, out} = 6$ and CN$_{\rm kp, inner}$ = 8, are distributed at the out and inner edges on the surfaces, defined as vertices \cite{liu09jpcc}; and CN$_{\rm kp, sub}$ = 10 at the subsurfaces.

Increasing ambient temperature, it accelerates the diffusion of atoms, and more subsurface atoms could diffuse towards the surface, giving rise to the increase of surface atoms number ($N_{\rm surf}$), as shown in Fig.~\ref{cnt}(b). It simultaneously triggers the surface reconstruction, leading to the change of local structure and surface energy for nanoporous. The evolution of surface rearrangement is shown in Fig.~\ref{cnt}(b). It is observed that surface reconstruction undergoes about four stages, denoted as I--IV. At the first stage (stage I, $T/T_{\rm M} < 0.44$), surfaces keep smooth and stable [600 K ($T/T_{\rm M}=0.44$), Fig.~\ref{sre}]. All of the atoms with different CN ($N_{\rm CN}$) are almost constant. Further to raise the temperature (0.44 $\le T/T_{\rm M} <$ 0.63, stage II), the kink sites, including edge of facets (kinked lines) and vertices (kinked points) first change, leading to an apparent decrease of $N_{\rm CN}$ for atoms at vertices with CN$_{\rm kp, out} = 6$, but a rise of $N_{\rm CN}$ for atoms at edge with CN$_{\rm kl, ed} = 7$. The other surface sites almost keep stable. Thus the surfaces remain smooth, except the transformation of kinked lines from the curve-shaped to some continuous-steps-shaped ones (800 K, Fig.~\ref{sre}). Then $N_{\rm CN}$ of edge atoms (CN$_{\rm kl, ed}$ = 7) decrease at stage III (0.63 $\le T <$ 0.81), giving rise to the increase of $N_{\rm CN}$ for the atoms at (111) facets (CN$_{\rm s(111)}$ = 9). The kinked lines become shorter and ``chaotic'', in contrast to the regular straight edges, causing the surface roughening [1050 K ($T/T_{\rm M} = 0.78$), Fig.~\ref{sre}]. As $T \ge$ 1100 K ($T/T_{\rm M} \ge 0.81$,stage IV), $N_{\rm CN}$ for atoms at edged lines and (111) facets decrease distinctly, accompanied by the substantial increase of $N_{\rm CN}$ for atoms CN = 11. Dense of ``chaotic'' and shorter kinked lines are distributed on the surfaces [1300 K ($T/T_{\rm M}=0.96$), Fig.~\ref{sre}], accelerating the surface roughening. Surface roughening then facilitates the nucleation and growth of vacancies in the subsurface and even the interior of crystals, which boost the inner disordering.

Surface melting begins as further to increase temperatures, and prefers to nucleate at the ``chaotic'' sites with dense of kinks [1350 K ($T/T_{\rm M} = 1.0$), Fig.~\ref{sre}], where the maximum ``disordering'' and free volume are presented, and the energy barrier to melt nucleation is the lowest, rather than smooth and ``ordered'' (111) facets at 1350 K. Melt propagates along the surface and then spreads into grain interiors by absorbing vacancies clusters or interstitial atoms near the inner faces. For other np-Cu samples with higher $\gamma$, it shows the similar surface reconstruction phenomena, although their melting temperatures are different.

\section{IV. Conclusion}
Our MD simulations demonstrate an apparent effect of specific surface area, on the collapse for a set of open-cell nanoporous Cu with the same porosities and shapes almost, during thermal annealing. Surface premelting dominates in their collapses, and surface premelting temperatures reduce linearly with increasing the specific surface area. The collapse mechanisms are different with different specific surface area. For np-Cu with the smaller specific surface area ($\gamma$ $\le$ 2.38 nm$^{-1}$), surface premelting is the cause to lead their collapse. While preferential surface premelting and following recrystallization predominate the pinch-off in np-Cu with $\gamma$ $>$ 2.38 nm$^{-1}$, where recrystallization arises at temperatures below supercooling ($T_{M-}$) owing to the instability and instance for liquid. Thermal-induced surface reconstruction prompts surface premelting by accelerating local ``disordering'' and ``chaotic'' in the surface, which are the preferred sites for surface premelting.

\section{acknowledgment}
This work was partially supported by the NSFs of China (Nos. 11672254, 51301066, and 51501063) and NSF of Hunan Province (Nos. 13JJ4071 and 14JJ2080). We would also like to acknowledge the support of the computation platform of the National SuperComputer Center in Changsha (NSCC).

\section{Appendix}
To describe the nucleation of melting in nanostructures, the Gibbs-Thomson equation, derived from classical nucleation theory (CNT) \cite{porter92b}, is used. For a solid spherical nanoparticle with the size of $d$, the change of Gibbs free energy during melting is expressed as \cite{kashchiev00b, wu16jmr}
\begin{equation}\label{gib1}
\Delta G = -\pi d^3\Delta G_{\rm V}/6 + \pi d^2 \sigma_{\rm sl},
\end{equation}
where $\sigma_{\rm sl}$ is the solid-liquid interfacial energy; $\Delta G_{\rm V}$ is Gibbs free energy difference per unit volume between solid and liquid phase, and
\begin{equation}\label{dg}
 \Delta G_{\rm V} = \left(1-T/T_{\rm M}\right)\Delta H_{\rm f}/V_{\rm s},
\end{equation}
here $\Delta H_f$ is the latent heat of fusion, and $V_{\rm s}$ is the molar volume of solid phase. When maximizing $\Delta G$, Gibbs-Thomson equation, describing the relation of particle size and corresponding melting temperature, is obtained \cite{jackson90jcp, jones74jms}
\begin{equation}\label{tm2}
 T = T_{\rm M}\left(1-\frac{\zeta}{d}\right),
\end{equation}
and
\begin{equation}\label{z1}
 \zeta = 4\sigma_{\rm sl}V_{\rm s}/\Delta H_f.
\end{equation}

For a nanoparticle, $\sigma_{\rm sl}$ is size dependent \cite{bai05prb, liang02jmsl}, and
\begin{equation}
\sigma_{\rm sl}(d) = \frac{2S_{\rm vib}(d)\Delta H_f(d)h}{3V_{\rm s}R}.
\end{equation}
Here $R$ is the ideal gas constant, $\sim$8.314 J mol$^{-1}$ K$^{-1}$; $h$ is the atomic diameter, $\sim$0.256 nm for Cu; and $S_{\rm vib}$ is the vibrational contribution of overall melting entropy of bulk crystals, which is a weak function of $d$, and could be ignored as a first-order approximation \cite{guisbiers09jpcc}, that $S_{\rm vib}$($d$) $\approx$ $S_{\rm vib}$($\infty$) $\approx$ 7.85 J mol$^{-1}$ K$^{-1}$ \cite{zhu09jpcc}. Substituting $\sigma_{\rm sl}$($d$) and $S_{\rm vib}$($d$) into Eq.~\ref{z1}, we have
\begin{equation}\label{z2}
 \zeta = 8hS_{\rm vib}(\infty)/3R.
\end{equation}
It reasonably matches the experimental data with $d\ge10$ nm \cite{castro90prb, dick02jacs} as the crystal retains its bulk values of $\sigma_{\rm sl}$, $\Delta H_f$ and $S_{\rm vib}$ \cite{jackson90jcp, peters98prb}. However, it fails for small-sized nanoparticles, adopting nonspherical shapes \cite{sun02sci, barnard09an, xia09acie} with a large $\gamma$. Here, a shape factor \cite{qi04mcp, qi16acr}, $\delta$=$A_{\rm NP}/A_{\rm SN}$, where $A_{\rm SN}$ is the surface area for a spherical nanoparticle and $A_{\rm NP}$ is the surface area of a nonspherical nanoparticle with the same volume as spherical nanoparticle, is used to describe the shape effect of nanoparticles. For a solid nonspherical particle with size of $d$ and shape factor of $\delta$, its surface area $A_{\rm solid} = \delta\pi d^2$. Combined with Eqs.~\ref{gib1}--\ref{z1}, the melting temperature of nonspherical nanoparticle is
\begin{equation}\label{z2}
 T = T_{\rm M}\left(1-\frac{\delta\zeta}{d}\right).
\end{equation}

The melting temperature of np-Cu with the identical sized nanopores, should be equal to that for the nanoparticle with the same size and shape. For a closed-cell np-Cu containing $N$ spherical nanopores with an identical size, $d$, we assume that $V_{\rm sp, unit}$ and $A_{\rm sp, unit}$ are the volume and surface area of a spherical nanopore, respectively. The system volume, solid volume and total surface area of sphere-shaped nanopores in the samples are denoted as $V_{\rm sys}$, $V_{\rm solid}$, and $A_{\rm sp}$, respectively. Here $NV_{\rm sp, unit}=V_{\rm sys}-V_{\rm solid}$, and $NA_{\rm sp, unit}=A_{\rm sp}$. It is noted that $A_{\rm sp} = A_{\rm solid}$, the surface area of solid, for the np-Cu with sphere-shaped nanopores. Consequently, the size of spherical nanopore is
\begin{equation} \label{foamd}
 d = \frac{6V_{\rm sp, unit}}{A_{\rm sp, unit}}=\frac{6(V_{\rm sys}-V_{\rm solid})}{A_{\rm sp}}.
\end{equation}
As $\gamma=S_{\rm solid}/V_{\rm solid}$, $\rho_r = V_{\rm solid}/V_{\rm sys}$, and $A_{\rm sp} = A_{\rm solid}$, Eq.~\ref{foamd} can be rewritten as
\begin{equation} \label{fd1}
 d = \mu/\gamma,
\end{equation}
where $\mu = 6(1-\rho_r)/\rho_r$.
For np-Cu containing $N$ nonspherical nanopores, $A_{\rm sp} < A_{\rm solid}=NA_{\rm np, unit}$, where $A_{\rm np, unit}$ is the surface area of a nonspherical nanopore. Then Eq.~\ref{fd1} can be rewritten as
\begin{equation}\label{fd2}
  d = \delta\mu/\gamma,
\end{equation}
here $\delta$ is the shape factor, and $\delta = A_{\rm solid}/A_{\rm sp}$. For an open-cell np-Cu, it can be considered to contain a nonspherical nanopore, whose size is $d = \sqrt[3]{6(V_{\rm sys}-V_{\rm solid})/\pi}$. Thus the melting temperature of open-cell np-Cu fomas can be obtained
\begin{equation}\label{fd3}
 T = T_{\rm M}\left(1-\zeta\gamma/\mu\right),
\end{equation}
indicating that the melt of nanoporous is the function of specific surface area ($\gamma$), liquid-solid interfacial energy ($\sigma_{\rm sl}$ in $\zeta$), and mass density ($\rho_r$ in $\mu$). For our constructed np-Cu, $\rho_r$ are almost constant, $\sim$ 58.20$\%$; and their sizes are almost the same, $d \approx 35.0$ nm $>$ 10 nm, implying $\sigma_{\rm sl}$ is the bulk values.

\bibliographystyle{apsrev}


\end{document}